\documentclass[floatfix,showpacs,prl,twocolumn]{revtex4}
\usepackage{eurosym}
\usepackage{graphicx}
\usepackage{graphics}
\usepackage{amssymb}
\usepackage{amsmath}
\usepackage{color}

\newcommand{\U}[1]{\sigma}

\begin{document}

\title{Field-controlled conical intersections in the vortex lattice of quasi
2D pure strongly type-II superconductors at high magnetic fields}
\author{T. Maniv}
\email{e-mail:maniv@tx.technion.ac.il}
\affiliation{Schulich Faculty of Chemistry, Technion-Israel Institute of Technology,
Haifa 32000, Israel}
\author{V. Zhuravlev}
\affiliation{Schulich Faculty of Chemistry, Technion-Israel Institute of Technology,
Haifa 32000, Israel}
\date{\today }

\begin{abstract}
It is shown that the Dirac fermion structures created in the middle of the
Landau bands in the vortex-lattice state of a pure 2D strongly type-II
superconductor at half-integer filling factors can be effectively controlled
by the external magnetic field. The resulting field-induced modulation of
the magneto-oscillations is shown to arise from Fermi-surface resonance
scattering in the vortex core regions. Possible observation of the predicted
effect in a quasi 2D organic superconductor is discussed.
\end{abstract}

\pacs{74.25.Ha, 74.25.Uv, 74.78.-w,74.70.Kn}
\maketitle

In a pure strongly type-II superconductor under a uniform magnetic field the
quasi particle spectrum is gapless in a broad field range below the upper
critical field $H_{c2}$ \cite{Dukan94},\cite{Dukan95},\cite{Maniv01}. In
this field range scattering of quasi particles by the vortex lattice
interferes with the Landau quantization of the electron motion perpendicular
to the magnetic field to form magnetic (Landau) Bloch's bands. The physical
picture is of an extended Bloch state which breaks magnetically into
localized cyclotron orbits \cite{Maniv01}. \ In pure 2D, or quasi 2D
superconductors, such as e.g. the organic charge transfer salt $\kappa
-\left( ET\right) _{2}Cu\left( SCN\right) _{2}$ \cite{Wosnitza96}, under a
magnetic field perpendicular to the easy conducting planes, the underlying
normal electron spectrum is fully quantized and the effect of the vortex
lattice is very pronounced. Furthermore, due to the suppressed energy
dispersion along the magnetic field direction and the particle-hole symmetry
inherent to the superconducting (SC) state, the quasi particle spectrum
exhibits peculiar features that are missing in the 3D case. For example, at
discrete magnetic field values where the chemical potential is located in
the middle of a Landau band, so that the underlying normal state spectrum
satisfies particle-hole symmetry, the calculated quasi-particle density of
states (DOS) has a Dirac Fermion structure \cite{Tesanovic-Sacramento98},
which reflects topological singularities at the vortex lattice cores.

In the present paper we reveal a physical mechanism which controls the Dirac
fermion structures created in the magnetic Brillouin zone (BZ) of the vortex
lattice of 2D strongly type-II superconductors at high magnetic fields, and
discuss possible experimental probes of their appearance. The ability to
create and control Dirac fermions just by varying an external parameter
(magnetic field in our case) is of great importance for future technological
applications (see, e.g. \cite{Kitaev03},\cite{Nayak08}). For the present
analysis we consider a model of a 2D electron system under a perpendicular
uniform magnetic field $\mathbf{H=}\left( 0,0,H\right) $, neglecting, for
the sake of simplicity, the Zeeman spin splitting and assuming a singlet, $s$%
-wave electron pairing. It should be noted that in a 2D or quasi 2D electron
system the condition of zero spin-splitting can be realized by tilting the
magnetic field direction with respect to the easy conducting planes \cite%
{Wosnitza96,Harrison12}.

The corresponding equations for the quasi particle states in the mean-field
approximation are the well known Bogoliubov de Gennes (BdG) equations in the
Landau orbitals representation \cite{Dukan94,Norman95}:

\begin{eqnarray}
\sum\limits_{n^{\prime }}\Delta _{n,n^{\prime }}\left( \mathbf{k}\right)
v_{n^{\prime }}\left( \mathbf{k}\right) &=&\left( \varepsilon -\xi
_{n}\right) u_{n}\left( \mathbf{k}\right) ,  \notag \\
\sum\limits_{n^{\prime }}\Delta _{n^{\prime },n}^{\ast }\left( \mathbf{k}%
\right) u_{n^{\prime }}\left( \mathbf{k}\right) &=&\left( \varepsilon +\xi
_{n}\right) v_{n}\left( \mathbf{k}\right) ,  \label{BdG}
\end{eqnarray}%
where the single-electron energy measured relative to the chemical potential 
$\mu $ is given by $\xi _{n}=\hbar \omega _{c}\left( n-n_{F}\right)
,n=0,1,2,...$, $n_{F}=\mu /\hbar \omega _{c}-1/2$, and $\omega
_{c}=eH/m^{\ast }c$ is the electronic cyclotron frequency.\ The matrix, $%
\Delta _{n,n^{\prime }}\left( \mathbf{k}\right) $, is diagonal in the
magnetic Brillouin zone, but non-diagonal in the Landau-level (LL) indices $%
n,n^{\prime }$. The pair potential $\Delta \left( \mathbf{r}\right) $
should,in principle,be determined self consistently with the eigenfunctions $%
v_{n}\left( \mathbf{k}\right) ,u_{n}\left( \mathbf{k}\right) $ \cite{Maniv01}%
. It will be very helpful to avoid the complexity involved in a fully
self-consistent approach by assuming $\Delta \left( \mathbf{r}\right) $ to
have the form of a vortex lattice, as will be elaborated below (see also SM2 
\cite{Supplem}). Since the Abrikosov vortex lattice shares with the self
consistent pair-potential in the lowest LL approximation \cite{Maniv01} the
feature of main interest here (i.e. the topological singularity of the
vortex-lattice cores) this is a reasonable assumption.

A very important information is encoded in these matrix elements\cite%
{Dukan94}: The zeros of the diagonal ground LL matrix-element $\Delta
_{0,0}\left( \mathbf{k}\right) $, are all of the first order and form a
lattice dual to the Abrikosov vortex lattice rotated by $90^{\circ }$. \
Matrix elements (diagonal as well as off-diagonal) with higher LL indices
have also zeros of higher orders. In the diagonal LL approximation, which is
valid sufficiently close to $H_{c2}$, and in the case where $n_{F}$
coincides with a LL index, $n=n_{F}$ (i.e. corresponding to particle-hole
symmetry in the normal state), one trivially solves the BdG equations to
find the quasi-particle energies: \ $\varepsilon _{\pm }\left( \mathbf{k}%
\right) =\pm \left\vert \Delta _{n,n}\left( \mathbf{k}\right) \right\vert $
, and the corresponding eigenstates: $\left( u_{n\pm }\left( \mathbf{k}%
\right) ,v_{n\pm }\left( \mathbf{k}\right) \right) =\left( \pm e^{i\phi
\left( \mathbf{k}\right) /2},e^{-i\phi \left( \mathbf{k}\right) /2}\right) /%
\sqrt{2}$ , with: $e^{i\phi \left( \mathbf{k}\right) }=\Delta _{n,n}\left( 
\mathbf{k}\right) /\left\vert \Delta _{n,n}\left( \mathbf{k}\right)
\right\vert $. \ Thus, near each vortex core, $k_{j}$, in the reciprocal
vortex-lattice \cite{Dukan94}, where $\left\vert \Delta _{n,n}\left( \mathbf{%
k}\right) \right\vert \rightarrow \eta _{n}\left\vert \mathbf{k-k}%
_{j}\right\vert \rightarrow 0$, the corresponding quasi-particle dispersion
relation exhibits a conical intersection of the $\pm $branches at the
chemical potential (zero energy), in close similarity to graphene Dirac cone
structure on a single honeycomb sub-lattice \cite{Katsnelson12},\cite%
{Rosenstein13}. \ Taking into account off-diagonal LL pairing, which is
crucial at magnetic fields even slightly away from $H_{c2}$, requires
numerical solution of Eq.(\ref{BdG}) with a great loss of physical insights.
However, the topological nature of the vortex cores singularity indicates
that significant fingerprints of this singularity should appear in the
dispersion relation under the influence of off-diagonal LL pairing. The
results for the quasi-particle density of states (DOS), shown in Fig.(1),
support this conjecture: Each broadened LL splits into two sub-bands due to
Andreev scattering of quasi particle-quasi hole in the vortex regions (see
e.g. \cite{Norman96}). The split sub-bands join to form straight-line 
\begin{figure}[tbp]
\label{fig1} \includegraphics[width=3in]{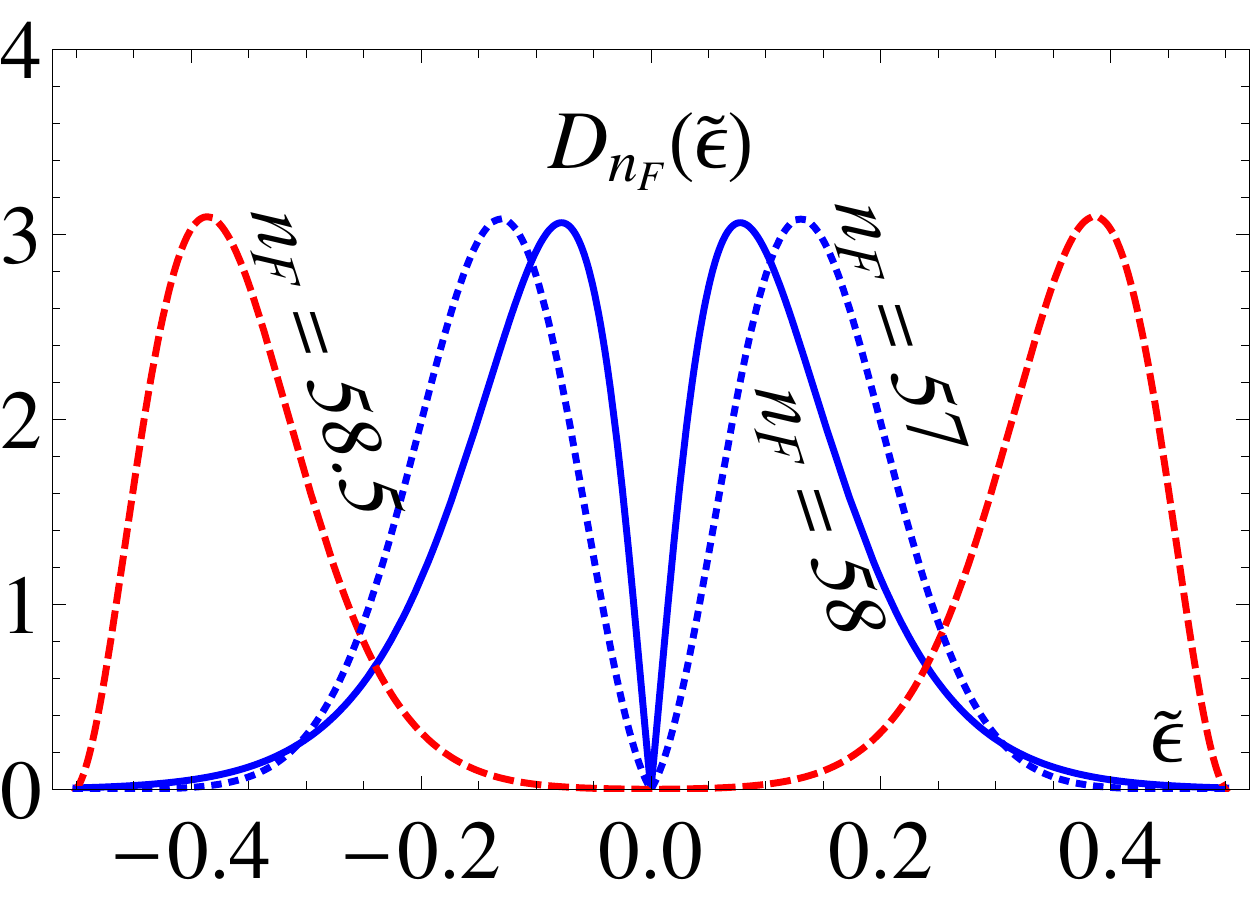}
\caption{{}Quasi-particle DOSs, as functions of $\widetilde{\protect%
\varepsilon }\equiv \protect\varepsilon /\hbar \protect\omega _{c}$,
calculated by solving the BdG equations (\protect\ref{BdG}) near the Fermi
energy ( $\protect\varepsilon =0$) for integers $n_{F}$ ($=57,58$) and a
half integer $n_{F}$ ($=58.5$) with $\Delta _{0}/\hbar \protect\omega _{c}=1$
. Note the sharp increase of the Dirac-shaped DOS slop from $n_{F}$ $=57$ to 
$58$. }
\end{figure}
intersection whenever the chemical potential is located in the middle of a
Landau band and the filling factor of the Landau levels, $\mu /\hbar \omega
_{c}=n_{F}+1/2$ , is half-integer ($n_{F}$ is integer). The linear vanishing
of the quasi-particle DOS at the Fermi energy (see Fig.1) is a consequence
of avoiding crossings, close to conical intersections, of quasi-particle
energy branches at many wave vectors in the 2D magnetic Brillouin zone (see
SM 1). At non-integer $n_{F}$, where the two (particle-hole) branches of the
quasi-particle energy in the diagonal approximation do not intersect, the
exact energy spectrum develops a gap between the two branches,$\pm
\left\vert \Delta _{n,n}\left( \mathbf{k}\right) \right\vert $, with a very
small number of states arising from off-diagonal LL pairing occupying the
'diagonal gap'. \ Fig.(1) illustrates the situation for a half integer $%
n_{F} $, where particle-hole symmetry is satisfied in both the normal and SC
states. \ For any other non-integer $n_{F}$ value this symmetry is obeyed in
the SC state, but not in the normal state. At integer values of $n_{F}$
normal-electron Landau tubes cross the (cylindrical) Fermi surface and the
oscillatory magnetization has maxima, whereas the minima of the oscillations
occur at the center of the cyclotron gaps (i.e. at half integers $n_{F}$ ).
The gaps developed in the quasi-particle spectrum in the vortex state at
half integers $n_{F}$ ensure that the minima of the oscillations also occur
at half-integer values of $n_{F}$. The envelope of the magneto-quantum
oscillations in the vortex state is therefore controlled by Dirac-shaped
quasi-particle DOS at integer values of $n_{F}$.

To reveal the physical mechanism controlling the field dependence of the DOS
one must complement the numerical approach of the BdG equations by resorting
to an analytical method that is sensitive to quasi particle (Andreev)
scattering in the vortex core regions. The exact Gorkov-Ginzburg-Landau
perturbation approach developed in Ref.(\cite{GGLex}) possesses the required
sensitivity. Following Ref.(\cite{GGLex}) the leading order term in the SC
thermodynamic potential, that is sensitive to vortex structure, i.e. the
quartic term, takes the form: $\Omega _{4}=N\left( k_{B}T/2\right) 
\widetilde{\Delta }_{0}^{4}\sum_{\nu =0}^{\infty }I_{4\nu }$, where $N$ is
the total number of flux lines threading the SC sample, $\widetilde{\Delta }%
_{0}=\left( \Delta _{0}/\hbar \omega _{c}\right) $ , and $I_{4\nu }$ is
written as a 4D 'temporal' integral$\ I_{4\nu
}=\prod\limits_{j=1}^{4}\int_{0}^{\infty }d\tau _{j}\left( \beta \left(
\gamma \right) /\alpha \right) e^{-\varpi _{\nu }\tau _{+}-in_{F}\tau _{-}}$%
. \ In this expression\ $\tau _{+}=\sum_{j=1}^{4}\tau _{j}$ , $\tau
_{-}=\sum_{j=1}^{4}\varepsilon _{j}\tau _{j}$, $\varepsilon _{j}\equiv
\left( -1\right) ^{j+1}$, $\alpha =\sum_{j=1}^{4}\alpha _{j}$, $\alpha
_{j}\equiv 1-e^{i\varepsilon _{j}\tau _{j}}$ , $\beta \left( \gamma \right)
=\sum_{\mathbf{G}}\left[ e^{-\theta _{-}\left\vert \mathbf{G}\right\vert
^{2}}/\left( 1+\gamma \right) +e^{-\theta _{+}\left\vert \mathbf{G}%
\right\vert ^{2}}/\left( 1-\gamma \right) \right] /2$, where $\gamma =\left(
\alpha _{2}\alpha _{4}-\alpha _{1}\alpha _{3}\right) /\alpha $ , $\theta
_{-}=\left( 1-\gamma \right) /\left( 1+\gamma \right) $, and $\theta
_{+}=\left( 1+\gamma \right) /\left( 1-\gamma \right) $. In $\beta \left(
\gamma \right) $ the summation is over the reciprocal vortex-lattice
vectors, $\mathbf{G}$ , measured in units of the inverse magnetic length $%
a_{H}^{-1}=\sqrt{eH/c\hbar }$, $\varpi _{\nu }\equiv \omega _{\nu }/\omega
_{c}$ , and$\ \ \omega _{\nu }=\left( 2\nu +1\right) \pi k_{B}T/\hbar ,\nu
=0,\pm 1,...$ is the Matsubara frequency at temperature $T$. The amplitude
of the SC order parameter, $\Delta _{0}^{2}=S^{-1}\int d^{2}\mathbf{r}%
\left\vert \Delta (\mathbf{r})\right\vert ^{2}$, with\textit{\ }$S=N\pi
a_{H}^{2}$, is treated as a variational parameter for minimizing the
thermodynamic potential $\Omega _{SC}\left( \Delta _{0}\right) .$The salient
features of $I_{4\nu }$ are: (1) The simple oscillatory factor $%
e^{-in_{F}\tau _{-}}$, revealing directly the Fourier transformed components
with respect to the dHvA frequency $F=n_{F}H$. \ (2) The vortex lattice
structure factor $\beta \left( \gamma \right) $- an extension of the well
known Abrikosov parameter to the high-field regime, which depends on the
electronic `temporal' variables, $\tau _{j}$\ , through a single composite
variable - $\gamma $, and on the vortex structure through the reciprocal
vortex lattice vectors $\mathbf{G}$. Remarkably, approaching the points $%
\gamma =+1,-1$ the divergence of the structure factor reflects singular
coupling of fermionic quasi particles to the vortex lattice. \ There are two
types of singular contributions: (a) where ( $1\pm \gamma $) vanishes in the
denominators of $\theta _{\pm }$, and (b) where ($1\pm \gamma $) vanishes in
the numerators. \ Case (a) corresponds to singular contribution from the
entire vortex lattice (i.e. from the single terms with $G=0$), whereas case
(b) corresponds to contributions from the entire reciprocal vortex lattice,
that is local in the direct vortex lattice. At the singular points, $\gamma
=+1,-1$, the exponential factor $e^{-in_{F}\tau _{-}}\rightarrow e^{-2\pi
inn_{F}}$, contributing only purely harmonic terms to the SC free energy in
the dHvA frequency $F=n_{F}H$.

Slightly away from the singular points that are local in the direct vortex
lattice, i.e. corresponding to many Umklapp scattering channels, there are
significant contributions to the SC free energy which deviate markedly from
harmonic behavior. They originate from G-vectors satisfying: $\left\vert 
\mathbf{G}\right\vert \approx 2\sqrt{2n_{F}}$, namely having length close to
the Fermi surface diameter. Furthermore, due to the incommensurability of
the large circular Fermi surface with the fine polygonal vortex lattice,
this \textit{Fermi surface resonance} condition yields erratic jumps of $%
I_{4\nu }$ as a function of $n_{F}$. The final result for the first harmonic
(i.e. $\tau _{-}=2\pi $ ) of the thermodynamic potential, $\Omega
_{SC}^{\left( 1h\right) }$, modulated by umklapp scattering effects in the
vortex lattice, up to fourth order in $\widetilde{\Delta }_{0}$, takes the
form: $\Omega _{SC}^{\left( 1h\right) }/\Omega _{N}^{\left( 1h\right)
}\simeq 1-\left( \pi ^{3/2}/\sqrt{n_{F}}\right) \widetilde{\Delta }_{0}^{2}+%
\left[ 1+w\left( n_{F}\right) \right] \left( \pi ^{3}/2n_{F}\right) 
\widetilde{\Delta }_{0}^{4}$, where $\Omega _{N}^{\left( 1h\right) }$ is the
corresponding normal state quantity, and $w\left( n_{F}\right) $, shown in
Fig.(2), represents highly anharmonic effects of the umklapp scattering by
the vortex lattice. \ The influence of vortex-lattice disorder is of
importance near $H_{c2}$ where random defects which pin flux lines, and/or
SC fluctuations, introduce disorder to the vortex lattice. The structure
factor, averaged over the disorder realizations in the white-noise limit,
reduces to its singular value, and the SC free energy up to fourth order in $%
\widetilde{\Delta }_{0}$, is purely harmonic, so that $\Omega _{SC}^{\left(
1h\right) }$ is obtained with $w\left( n_{F}\right) \rightarrow 0$, i.e.
very close to the well known Maki-Stephen expression \cite{Maki91,Stephen92}%
, as expanded to the same order in $\Delta _{0}$.

The great advantage of the perturbation approach just described is in the
ability to derive analytical expressions, at least for the leading terms,
with sensitivity to the Andreev scattering in the vortex core regions. On
the other hand, at any order of the perturbation expansion the expected
broadening effect of the LL is absent. To see whether the predicted erratic
oscillation effect survives this broadening we have calculated the
quasi-particle DOS at various integer values of $n_{F}$ by numerically
solving Eqs.(\ref{BdG}). The quantum oscillation (QO) amplitude obtained
from the resulting DOS, $\mathcal{D}_{n_{F}}\left( \varepsilon \right) $, by
means of the expression: $D\left( n_{F},T\right) \approx \left( \frac{%
m^{\ast }}{2\hbar ^{2}}\right) \int d\varepsilon \frac{\mathcal{D}%
_{n_{F}}\left( \varepsilon \right) X}{\cosh ^{2}\left( \varepsilon X\right) }
$, derived in the low temperature limit, $X\equiv \frac{\hbar \omega _{c}}{%
2k_{B}T}\gg 1$, of the well-known formula for the thermodynamic DOS \cite%
{Katsnelson12}, is compared in Fig.(2) with the oscillatory modulation
function $w\left( n_{F}\right) $. The good 
\begin{figure}[tbp]
%\caption{Top panel: Calculated thermodynamic}\label{fig3} %
\includegraphics[width=3.1in]{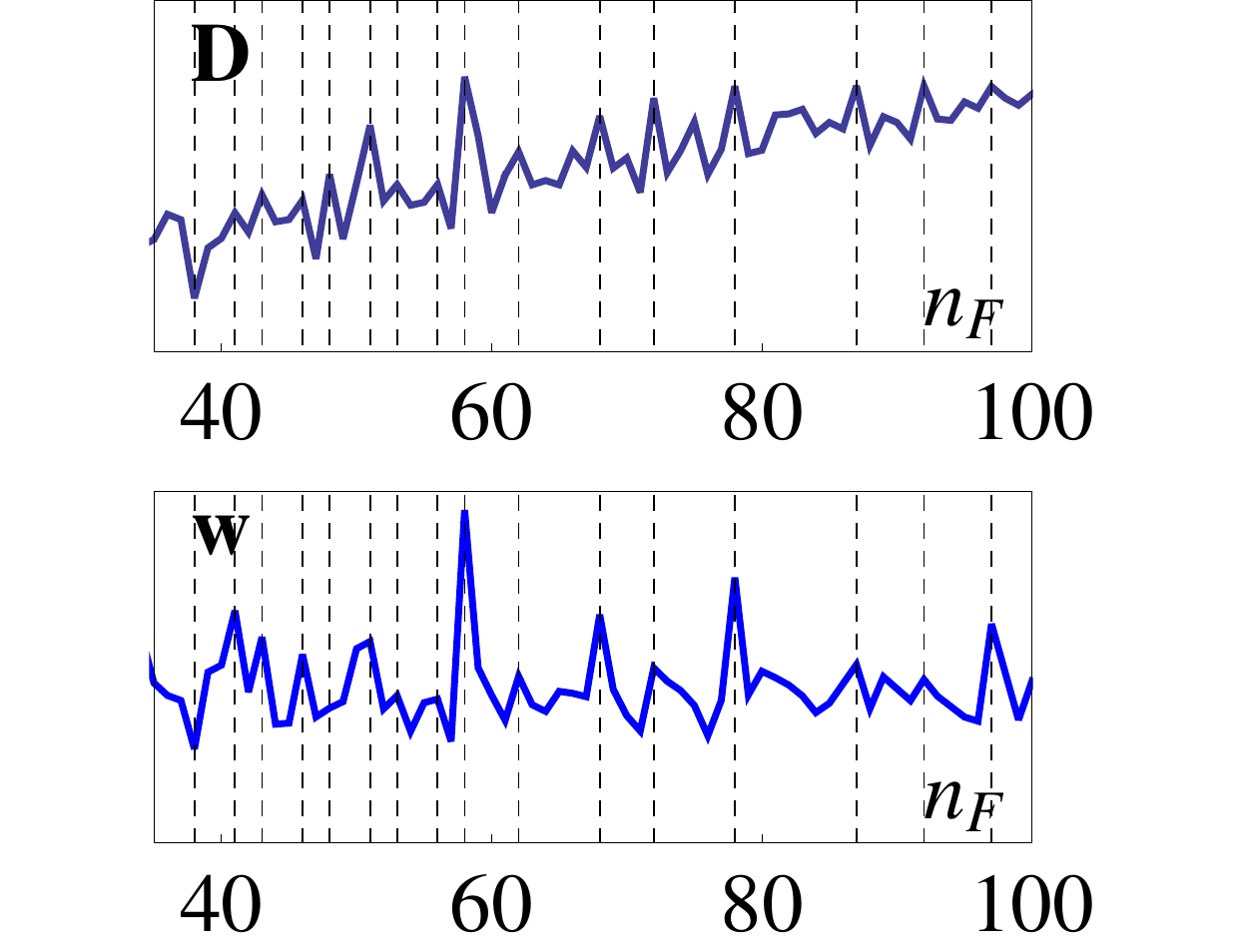}
\caption{{}Top panel: Calculated thermodynamic DOS \protect\cite%
{Katsnelson12}, as a function of integer $n_{F}$, obtained at $\hbar \protect%
\omega _{c}/2k_{B}T=10$, from the BdG equations. Bottom panel: QO amplitude
calculated at the same $n_{F}$ values using perturbation GGL theory. Note
the pronounced jump from $n_{F}=57$ to $58$ which correlates with a sharp
increase of the Dirac-shape slop shown in Fig.(1). }
\end{figure}
agreement between major features in the modulated envelopes of oscillations
obtained in these calculations confirms the conjectured relation between the
enhancement of the DOS slop at zero energy and the Fermi surface
resonance-scattering through the vortex lattice cores. The most pronounced
feature appearing quite similarly in both calculations is the sharp rise of
the QO amplitude seen in Fig.2 in going from $n_{F}=57$\ to $58$, which is
correlated with a sharp increase in the slop of the DOS at $\varepsilon =0$
(see Fig.1). The corresponding dispersion relation for $n_{F}=57$ (see SM1 
\cite{Supplem}) preserves the ideal conical intersection at the vortex core
which characterizes the diagonal LL branches $\pm \left\vert \Delta
_{n,n}\left( \mathbf{k}\right) \right\vert $. Note, however, that the
Dirac-shaped DOS appearing at integer $n_{F}$ values is determined by the
entire landscape of avoiding crossings close to conical intersections
appearing in the 2D BZ rather than by a single Dirac cone at the vortex core
(see SM1 \cite{Supplem}).

Experimental observation of the predicted effect is expected in strongly
type-II, layered superconductors in which magneto- quantum oscillations can
be observed in the mixed state \cite{Janssen98,Maniv01}. The only relevant
example known so far is $\kappa -\left( ET\right) _{2}Cu\left( SCN\right)
_{2}$, which was studied rather intensively by several groups\cite{van der
Wel95},\cite{Sasaki98}. Sufficiently small interlayer transfer integral
(i.e. about $0.04$ meV\cite{Singleton02}) on the scale of the cyclotron
energy (which is about $0.13$ meV at $H=4$ T, see below), ensuring 2D
electron dynamics, and clear dHvA oscillations in the SC mixed state with
significant SC-induced extra damping, have been reported for this material.
A modified version of the Maki-Stephen relaxation time approximation \cite%
{Maki91,Stephen92}, which takes into account the effect of SC fluctuations
in the vortex liquid state\cite{Maniv01} has shown a very good quantitative
agreement with the observed data \cite{Maniv01},\cite{Sasaki03}. The best
fitting value of the mean-field $H_{c2}\left( T\rightarrow 0\right) $
obtained in this analysis was $4.7$ T, which is consistent with the H-T
phase diagram derived in Ref.(\cite{Sasaki98}). However, observation of the
predicted Dirac Fermions fingerprint on the dHvA oscillation depends on
whether freezing of the vortex liquid occurs in the field range where
superconductivity and dHvA oscillations coexist (i.e. for $H\sim 4.2-4.7$ $T$%
). Ignoring possible quantum fluctuation effects this can occur at
sufficiently low temperatures where thermal fluctuations are suppressed. The
data reported in Ref.(\cite{van der Wel95}) (measured at $T=20$ mK) seems to
indicate that an ordered 2D vortex lattice indeed appears in the field range 
$H\approx 4.2-4.4$ $T$. To substantiate this statement we have calculated
the quasi-particle DOS using the BdG equations for a 2D vortex lattice with
the field-dependent order-parameter amplitude calculated in the mean-field
approximation (which is a good approximation in the field range investigated 
\cite{Maniv01}) and with parameters adopted from Ref.\cite{van der Wel95}
and from the best fit reported in Ref.(\cite{Maniv01}) (see SM2-\cite%
{Supplem}). The thermodynamic DOS oscillation was calculated by first
convoluting the quasi-particle DOSs with a Lorenzian of width $\gamma $
determined by the measured Dingle temperature in the normal state to take
into account the effect of atomic-lattice disorder. In the relevant field
range the order-parameter amplitude is large (i.e. $\widetilde{\Delta }%
_{0}\sim 3$) so that the Landau bands fill essentially the entire cyclotron
`gaps' (the broad-bands region). Two pairs of DOSs calculated at $%
n_{F}=155,155.5$ and $n_{F}=161,161.5$, bordering the field range of
interest, are shown in Fig.(3). In the effective energy interval ($2\gamma $%
) around the Fermi surface which dominates the QO, the DOS reduction in
going from $n_{F}=155$ to $155.5$ is much larger than in going from $161$ to 
$161.5$, though in both cases the bands essentially fill the entire
cyclotron energy interval. Thus, in the region of broad Landau bands, i.e.
for $n_{F}$ values above about $n_{F}=155$, the field modulation of the
thermodynamic DOS is controlled by the quasi-particle scattering
trajectories crossing vortex-lattice cores, rather than by the usual
crossing of Landau tubes through the Fermi surface. The resulting calculated
QO pattern, shown as inset in Fig.(4) as a function of $1/H$, exhibits a
crossover from regular oscillations, with a\ monotonically decreasing
amplitude, to a 'irregularly' modulated pattern, reflecting the sharp
field-modulation of the\ quasi-particle DOS under vortex-lattice cores
scattering. The calculation reproduces reasonably well a crossover of the
same type, starting at about $H=4.35$ \ T ($n_{F}=156$)
in the experimental data presented in Fig.(4).

\begin{figure}[tbp]
\label{fig4} \includegraphics[width=3.3in]{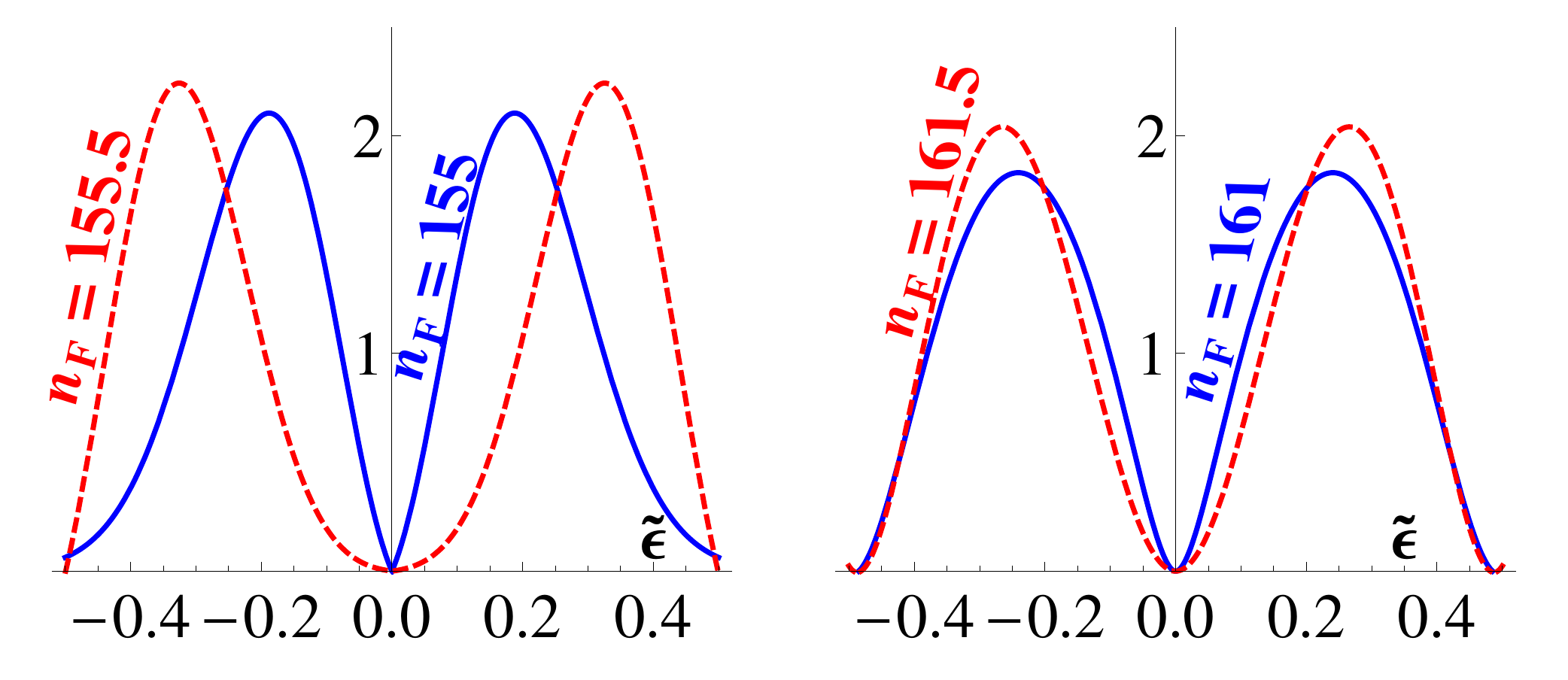}
\caption{{}DOSs, as functions of $\widetilde{\protect\varepsilon }\equiv 
\protect\varepsilon /\hbar \protect\omega _{c}$, calculated by solving the
BdG equations \protect\ref{BdG} in the broad-band region ($\widetilde{\Delta 
}_{0}\sim 3$) at $n_{F}=155$ (left panel, solid line) and $155.5$ (dashed
line), and at $n_{F}=161$ (right panel, solid line) and $161.5$ (dashed
line).}
\end{figure}
Some readers might argue that the sharp features seen in the experiment are
just noise. It should be stressed, however, that the magnitudes of the
calculated features (see, e.g. the sharp changes of the DOS with $n_{F}$,
and their fingerprints on the magneto-oscillations shown in SM2 \cite%
{Supplem} between $n_{F}=155$\ and $157$) are seen to be comparable to those
seen in the experimental data (Fig.(4)). It is\ therefore plausible that the
experimentally observed features arise from the predicted effect.

It should be also noted that a similar calculation done for the hexagonal
vortex-lattice model (not shown) has yielded in this broad-bands region
smoother features as compared to the square-lattice calculation. \ In the
narrow Landau bands region sharp features essentially similar (but different
in their details) to those shown in Figs.1,2 were obtained for the hexagonal
lattice model.

\begin{figure}[tbp]
\label{fig5} \includegraphics[width=3.5in]{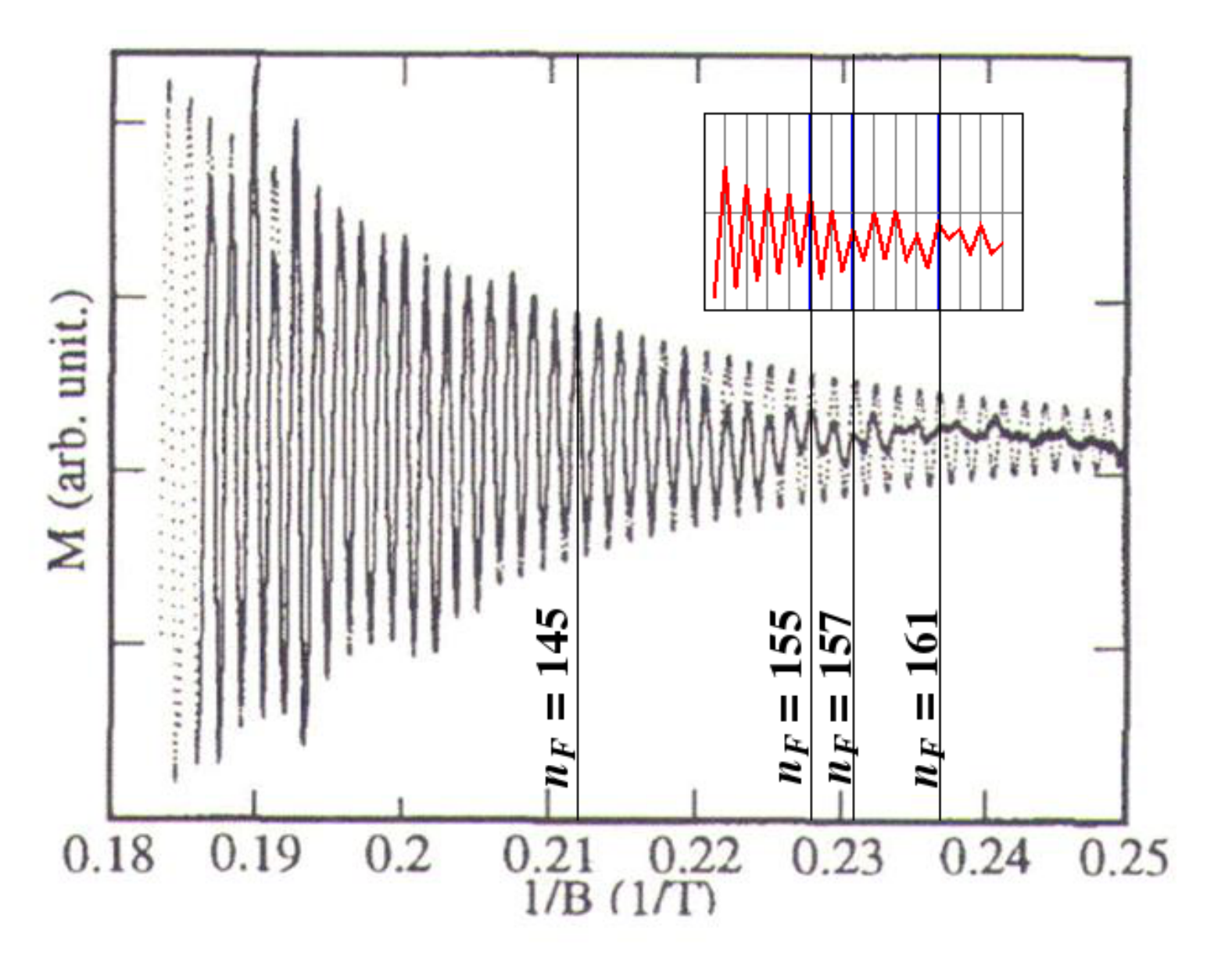}
\caption{dHvA oscillation above and below mean-field $H_{c2}\approx 4.7$ $T$
($n_{F}=145$) at $T=20$ mK reported in Ref.\protect\cite{van der Wel95}. The
dashed line extrapolates the normal state oscillation to the SC region. The
inset represents calculated QO obtained by solving the BdG equations for a
square vortex lattice as described in the text. \ Note the same scale on
orizontal axes ($n_{F}$) for the main figure and the inset. \ }
\end{figure}

In conclusion, we have revealed a fundamental relationship between Fermi
surface resonance quasi-particle scatterings through vortex lattice cores
and formation of pronounced Dirac Fermion structures in the reciprocal
vortex lattice of a 2D strongly type-II superconductor at high magnetic
fields. The predicted effect can be detected by finely tuning the external
magnetic field through resonant Andreev scattering channels which sharply
modulate the quasi-particle density of states. The effect is shown to leave
an observable fingerprint on magneto-quantum oscillations in the
vortex-lattice state of a quasi 2D organic superconductor and could be
directly detected in future scanning tunnelling spectroscopy measurements.
Since vortex-lattice disorder is expected to suppress the effect \cite{GGLex}
its observation could be used to identify the freezing transition into the
vortex-lattice phase.

%\begin{acknowledgments}
This research was supported by E. and J. Bishop research fund at Technion,
and by EuroMagNET under the EU contract No.\ 228043. V.Z. acknowledges the
support of the Israel Science Foundation by Grant No. 249/10 
%\end{acknowledgments}

\end{document}

\title{Field-controlled conical intersections in the vortex lattice of quasi
2D pure strongly type-II superconductors at high magnetic fields:
Supplemental material}
\author{T. Maniv and V. Zhuravlev}
\affiliation{Schulich Faculty of Chemistry, Technion-Israel Institute of Technology,
Haifa 32000, Israel}
\date{\today}
\maketitle

\section*{Supplemental material 1: Landau band-structure in the Brillouin
zone and Dirac-shaped density of states at half integer filling factors}

\begin{figure}[h]
\includegraphics[width=2.5in]{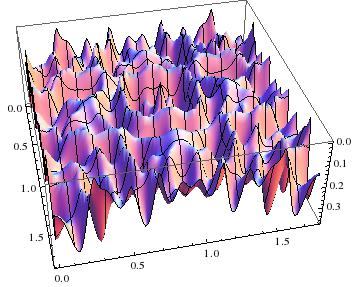}\hspace{.8cm}%
\includegraphics[width=2.5in]{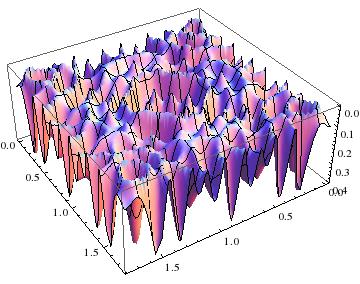} \newline
\includegraphics[width=2.5in]{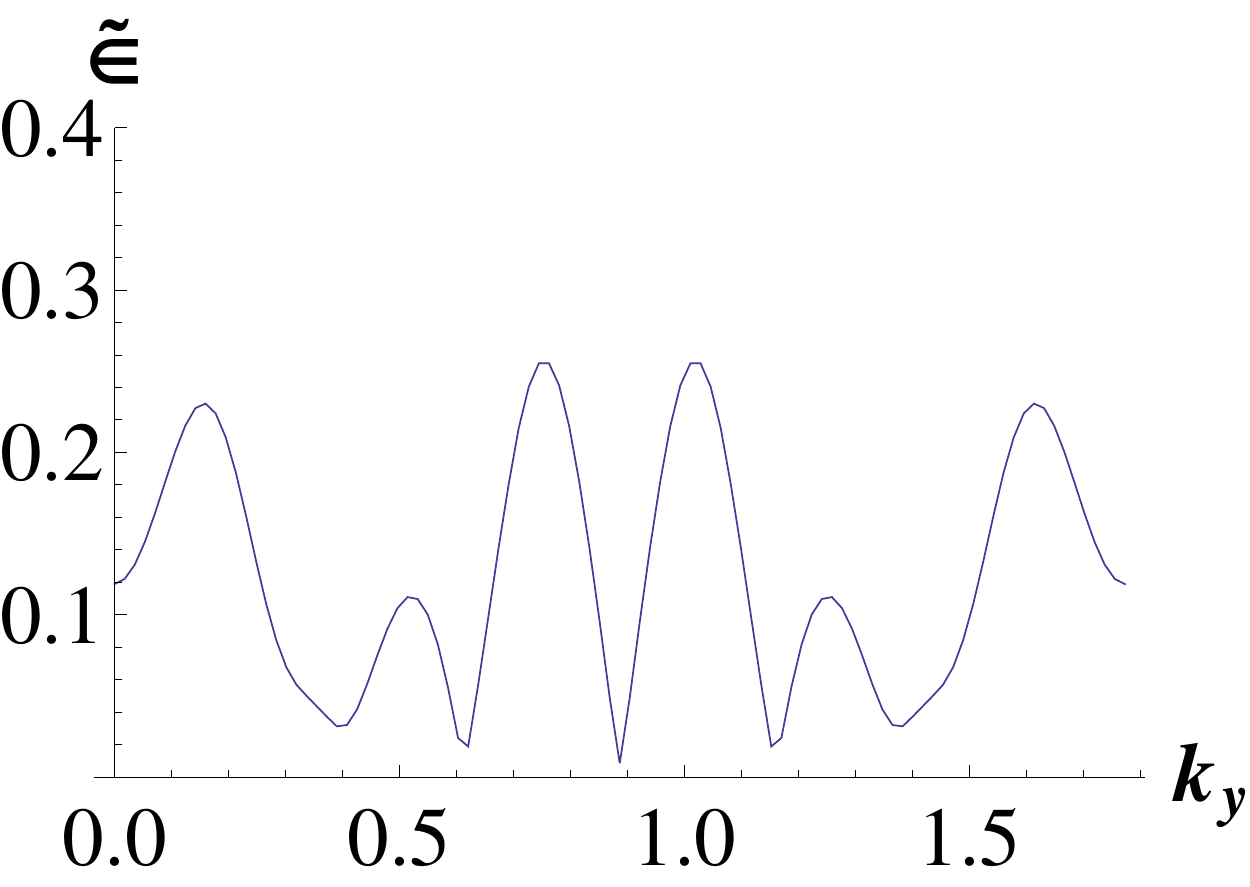}\hspace{.8cm}%
\includegraphics[width=2.5in]{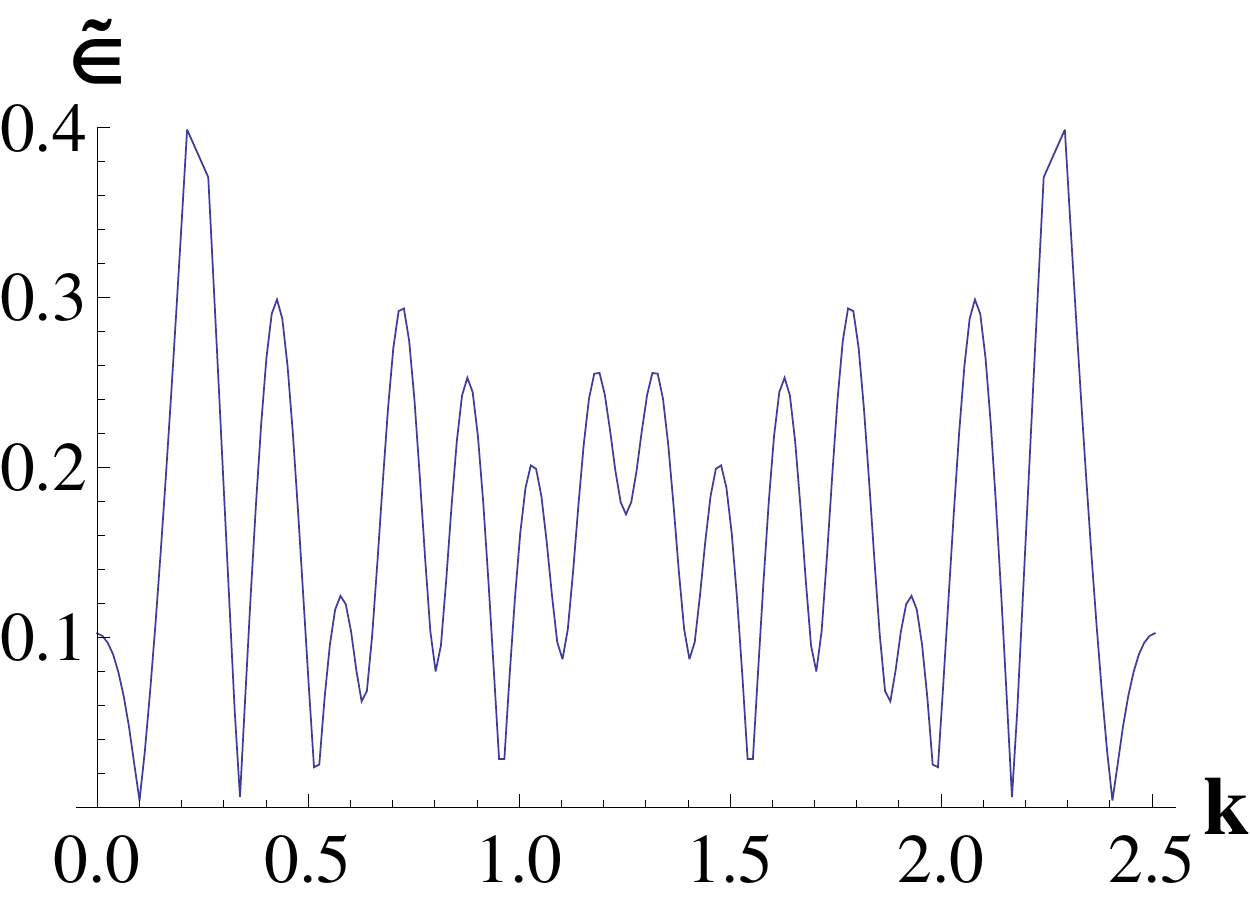}
\caption{Top panels: Landau band-structures for $n_{F}=57$ (left) and $%
n_{F}=58$ (right), calculated for the square Abrikosov vortex lattice in the
entire 2D BZ (note the inverted energy axes). Bottom panels: 2D plots along
symmety lines passing through the vortex core: For $n_{F}=57$ along the $%
k_{y}$-axis (left panel), whereas for $n_{F}=58$ along the diagonal axis
(right panel). \ The amplitude of the pair-potential was selected to be: $%
\Delta _{0}(H)/\hbar \protect\omega _{c}=1$.}
\label{SM11}
\end{figure}

\begin{figure}[th]
\includegraphics[width=4in]{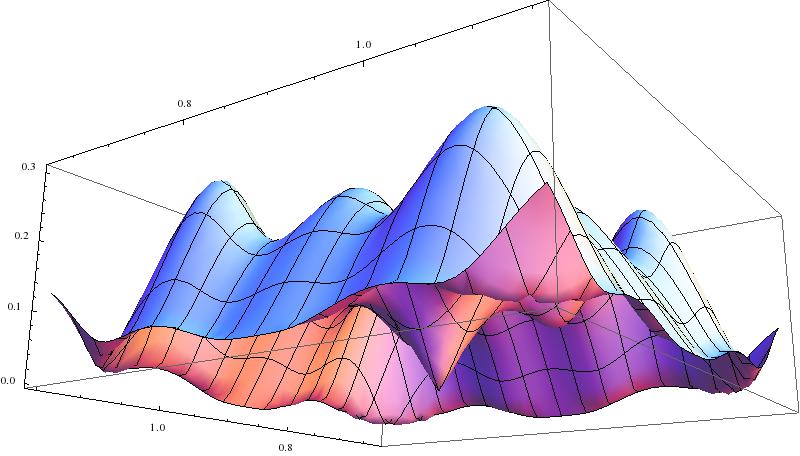}
\caption{The energy dispersion in a restricted range around the BZ center
(reciprocal vortex-lattice core position) for $n_{F}=57$, showing an
extension of the central part of the top-left panel in Fig.(1) with a clear
Dirac cone at the vortex core position. }
\label{SM12}
\end{figure}

In this supplemental material we illustrate how the complex Landau band
structures obtained from the BdG equations (Eq.(1) in the main text) for
relatively large integer values of $n_{F}$ transform into the simple
Dirac-shape density of states (DOS) functions, shown e.g. in Fig.(1) of the
main text. We also show here how the landscape characterizing such a band
structure can be controlled by varying the Landau-level filling factor, $%
F/H+1/2=n_{F}+1/2$, e.g. through the magnetic field intensity $H$. This is
illustrated here for the step from $n_{F}=57$ to $58$ where the most
pronounced jump in the quantum-oscillation amplitude is seen in Fig.(2) of
the main text. Starting with a 3D plot of the Landau band-structure in the
entire BZ for $n_{F}=58$ shown in Fig.(\ref{SM11}) (top-right panel) it is
seen to have a very rich landscape decorated with many avoiding crossings
close to conical intersections at zero energy (mostly with non-circular
directrices), which reflect the influence of the vortex-lattice cores
singularity. The linear energy dependence of the resulting DOS function near
zero energy, shown in Fig.(1) of the main text, is determined by integrating
over all these 'conical intersections'. Note, however, that at the vortex
core position the band structure for $n_{F}=58$ has no Dirac cone but a
crater at the top of a large peak (Fig.(\ref{SM11}) bottom-right panel).
Upon varying $n_{F}$ from $58$ to $57$ the high energy states contributing
to this peak are shifted away from the vortex core, clearing the landscape
for the Dirac cone at the vortex core to be visible (see Fig.(\ref{SM11})
left panels and Fig.(\ref{SM12}) ).

\section*{Supplemental material 2: Comparison with experimental dHvA
oscillations data}

To compare our theory with the available low temperature ($20$ mK)
experimental data of dHvA oscillations in the vortex state of the quasi 2D
organic superconductor $\kappa -\left( ET\right) _{2}Cu\left( SCN\right)
_{2} $ \cite{van der Wel95} we use the Abrikosov lattice form for the pair
potential \cite{Maniv01}: 
\begin{eqnarray}
\Delta \left( \mathbf{r}\right) &=&\left( \frac{2\pi }{a_{x}^{2}}\right)
^{1/4}\Delta _{0}e^{ixy}\sum\limits_{k}e^{-i\theta k^{2}+iq_{k}x-\left(
y+q_{k}/2\right) ^{2}},  \label{Delta} \\
q_{k} &=&\frac{2\pi k}{a_{x}},k=0,\pm 1,...,a_{x}^{2}=\pi /\sqrt{1-\left(
\theta /\pi \right) ^{2}}  \notag
\end{eqnarray}%
where the coordinates $\mathbf{r}$ are measured in units of the magnetic
length $a_{H}=\sqrt{c\hbar /eH}$ and $a_{x}$ is the lattice constant along
the main principal axis. In Eq.(\ref{Delta}) $\theta =0$ corresponds to the
square lattice whereas $\theta =\pi /2$ for the triangular lattice . For the
amplitude $\Delta _{0}$ we use the simple mean-field (BCS) form:%
\begin{equation}
\frac{\Delta _{0}\left( n_{F}\right) }{\hbar \omega _{c}}=1.31\left( \frac{%
T_{c}\left[ K\right] }{H\left[ T\right] }\right) \left( \frac{m_{c}^{\ast }}{%
m_{0}}\right) n_{F}\left( 1-\frac{F/H_{c2}}{n_{F}}\right)  \label{Delta0}
\end{equation}%
where $m_{c}^{\ast }$ is the cyclotron effective mass, $m_{0}$ the free
electron mass and $F=n_{F}H$ is the dHvA frequency. For the detected signal
in the vortex state \cite{van der Wel95} $F=680$ T, $\left( \frac{%
m_{c}^{\ast }}{m_{0}}\right) =3.5$ , and mean-field $H_{c2}$ which best fits
the extra damping in the vortex liquid state is $4.7$ T\cite{Maniv01},\cite%
{Sasaki03}.

\begin{figure}[t]
\label{fig1} \includegraphics[width=4in]{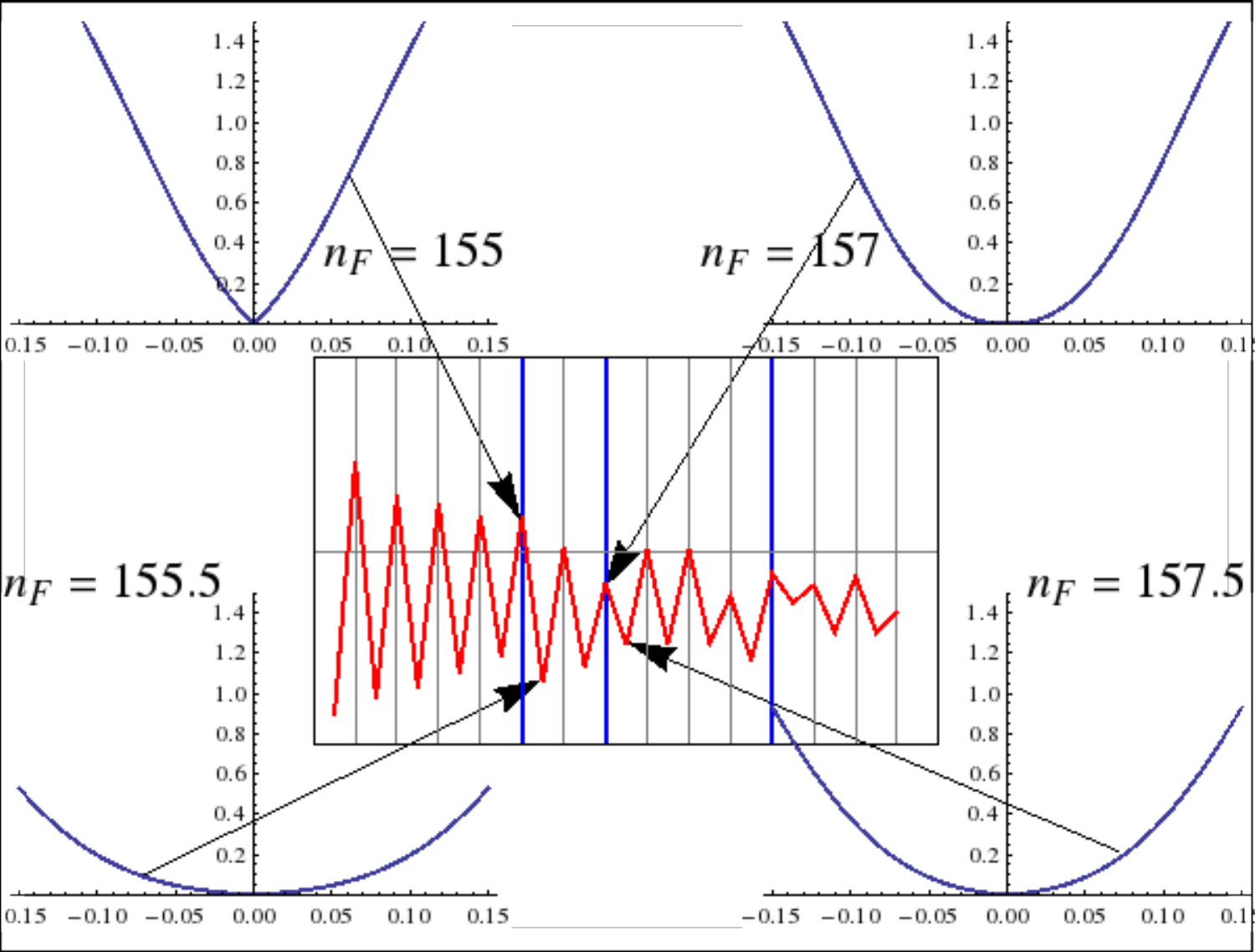}
\caption{Calculated $D\left( n_{F},T;\protect\gamma \right) $ in the range $%
n_{F}=151-164$ ($H=4.5-4.15$ T) (Central panel), showing sharp amplitude
modulation between $n_{F}=155$ and $n_{F}=157$ associated with the
suppression of the Dirac-shaped DOS shown in the upper-left panel (i.e. for $%
n_{F}=155$) upon moving to $n_{F}=157$ (upper-right panel).}
\label{SM2}
\end{figure}

The quasi-particle DOS,$\mathcal{D}_{n_{F}}\left( \varepsilon \right) $,
obtained from the solutions of Eq.(1) in the main text at various values of $%
n_{F}$ is convoluted with a Lorenzian of width $\gamma =2k_{B}\pi T_{D}$,
determined by the measured Dingle temperature $T_{D}$ in the normal state,
to take into account the effect of atomic-lattice disorder in the simple
relaxation time approximation, i.e.: 
\begin{equation*}
\mathcal{D}_{n_{F}}^{\gamma }\left( \varepsilon \right) =\frac{1}{\pi }\int
d\varepsilon ^{\prime }\mathcal{D}_{n_{F}}\left( \varepsilon ^{\prime
}\right) \frac{\widetilde{\gamma }}{\left( \varepsilon ^{\prime
}-\varepsilon \right) ^{2}+\widetilde{\gamma }^{2}},\widetilde{\gamma }%
\equiv \frac{\gamma }{\hbar \omega _{c}}
\end{equation*}

The thermodynamic DOS (or quantum capacitance) oscillation is finally
calculated by means of the expression:

\begin{equation}
D\left( n_{F},T;\gamma \right) \approx \left( \frac{m^{\ast }}{2\hbar ^{2}}%
\right) \int d\varepsilon \frac{X}{\cosh ^{2}\left( \varepsilon X\right) }%
\mathcal{D}_{n_{F}}^{\gamma }\left( \varepsilon \right)  \label{QC}
\end{equation}%
derived in the low temperature limit, $X\equiv \frac{\hbar \omega _{c}}{%
2k_{B}T}\gg 1$, of the well-known formula for the thermodynamic DOS \cite%
{Katsnelson12}. The value of $T_{D}$ determined from the normal state dHvA
oscillation damping \cite{Sasaki98} yields $\gamma =0.28$ K , that is $%
\widetilde{\gamma }\approx 0.16$ at $H=4.5$ T. \ \ Using the set of
parameters, described above, in Eq.(\ref{QC}), the resulting QO as a
function of integer and half-integer values of $n_{F}$ is shown in Fig.(\ref%
{SM2}). \ DOS plots in the effective quasi-particle energy range of $-%
\widetilde{\gamma }\lesssim \varepsilon \lesssim \widetilde{\gamma }$ for $%
n_{F}=155$,$155.5$ are compared there to similar plots for $n_{F}=157$, $%
157.5$. \ The sharp reduction of the DOS in going from $n_{F}=155$ to $157$
and to $155.5$ and the moderate reduction in going from $n_{F}=157$ to $%
157.5 $ are the origins of the sharp drop of the oscillation amplitude seen
between these points.